# Max-SINR Receiver for HMCT Systems over Non-Stationary Doubly Dispersive Channel

*Kui Xu, Youyun Xu, Dongmei Zhang, Wenfeng Ma*

College of Communications Engineering, PLA University of Science and Technology, Nanjing 210007, China.

Email:lgdxxukui@126.com

## Abstract

In this paper, a maximizing Signal-to-Interference plus-Noise Ratio (Max-SINR) receiver for Hexagonal Multicarrier Transmission (HMCT) system over non-stationary doubly dispersive (NSDD) channel is proposed. The closed-form timing offset expression of the prototype pulse for the proposed Max-SINR HMCT receiver over NSDD channel is derived. Simulation results show that the proposed Max-SINR receiver outperforms traditional projection scheme and obtains an approximation to the theoretical upper bound SINR performance within all the local stationarity regions (LSRs). Meanwhile, the SINR performance of the proposed Max-SINR HMCT receiver is robust to the estimation error between the estimated value and the real value of root mean square (RMS) delay spread.

## 1. Introduction

With the development of mobile communication systems such as Long Term Evolution (LTE), WiMax and International Mobile Telecommunications (IMT)-Advanced, broadband wireless multimedia services will be prevalent in the near future. Meanwhile, high speed vehicles such as airplanes and high speed trains play an increasingly important role in people's lives, since they provide a relatively stable and spacious environment for long distance travel. As a result, there will be a strong demand for broadband wireless communications (BWC) for high speed vehicles to provide information and on board entertainment services to passengers.

The key features of wireless channels in high speed vehicles communication systems are non-stationarity and doubly dispersive (DD) [1,2]. The existing orthogonal frequency division multiplexing (OFDM) based BWC systems can prevent inter-symbol interference (ISI) caused by multipath channel. Meanwhile, the time variations of the channel during one OFDM symbol duration destroy the orthogonality of different subcarriers, and result in power leakage among subcarriers, known as inter-carrier interference (ICI). In order to overcome the above drawbacks of OFDM system, several pulse-shaping OFDM and multi-carrier modulation (MCM) systems were proposed [3,4].

It is shown in [5–8] that signal transmission through a rectangular lattice is suboptimal for DD channel. By using results from sphere covering theory, the authors in [5–10] have demonstrated that hexagonal multi-carrier transmission (HMCT) system obtains lower energy perturbation than OFDM system, hence outperforms OFDM system from the robustness against channel dispersion point of view.

It is shown in [11,12] that the optimum sampling time of wireless communication systems over DD channel depends on the power distribution of the channel profiles, and that zero timing offset does not always yield the best system performance. In our previous work [9,10], we have analyzed the system SINR performance and presented that a traditional HMCT receiver proposed in [7,8] using a zero timing-offset prototype pulse is a suboptimal approach in the view of SINR [20-23].

In this paper, a novel receiver based on Max-SINR criterion for HMCT system over non-stationary DD (NSDD) channel is proposed. Theoretical analyses and simulation results show that the proposed optimal Max-SINR receiver outperforms traditional projection receiver and obtains an approximation to the theoretical upper bound SINR performance. Meanwhile, the proposed scheme is robust to the estimation error between the estimated value and the real value of root mean square (RMS).

## 2. System Model

In HMCT systems, the transmitted baseband signal can be expressed as [7]

$$x(t) = \sum_m \sum_n c_{m,2n} g(t-mT) e^{j2\pi nFt} + \sum_m \sum_n c_{m,2n+1} g(t-mT-T/2) e^{j2\pi(nF+F/2)t} \quad (1)$$

where $T$ and $F$ denote the symbol period and the subcarrier separation, respectively; $c_{m,2n}$ denotes the user data, which is assumed to be taken from a specific signal constellation and independent and identically distributed (i.i.d.) with zero mean and average power $\sigma_c^2$; $m \in \mathcal{M}$ and $n \in \mathcal{N}$ are the position indices in the TF plane; $\mathcal{M}$ and $\mathcal{N}$ denote the sets from which $m$, $n$ can be taken, with cardinalities $M$ and $N$, respectively. The prototype pulse $g(t)$ is the Gaussian window $g(t)=(2/\sigma)^{1/4}e^{-(\pi/\sigma)t^2}$, and $\sigma$ is the parameter controlling the energy distribution in the time and frequency directions. The ambiguity function of Gaussian pulse is defined by $A_g(\tau,\upsilon)=\int_{-\infty}^{\infty}g(t)g^*(t-\tau)e^{-j2\pi\upsilon t}dt$, where $(.)^*$ denotes the complex conjugate.

The hexagonal lattice can be expressed as the disjoint union of a rectangular sublattice $V_{rect1}$ and its coset $V_{rect2}$. The transmitted signal in (1) can be rewritten as $x(t)=\sum_i\sum_m\sum_n c_{m,n}^i g_{m,n}^i(t)$, $i=1,2$, $c_{m,n}^1$ and $c_{m,n}^2$ represent the symbols coming from $V_{rect1}$ and $V_{rect2}$, respectively. $g_{m,n}^i(t)=g(t-mT-\lfloor i/2\rfloor T)e^{-j2\pi(nF+\lfloor i/2\rfloor F)t}$ is the transmitted pulse generated by the prototype pulse $g(t)$, $\lfloor\cdot\rfloor$ denotes the floor function. .

The input-output relation of an NSDD channel $\mathbf{H}$ is [13,14]

$$(\mathbf{H}x)(t)=\int h(t,\tau)x(t-\tau)d\tau \qquad (2)$$

where is the impulse response of $\mathbf{H}$. Alternative channel descriptions are the time-varying transfer function, defined as $L_{\mathbf{H}}(t,f)=\int h(t,\tau)e^{-j2\pi f\tau}d\tau$ and the (delay-Doppler) spreading function, given by $S_{\mathbf{H}}(\tau,\upsilon)=\int h(t,\tau)e^{-j2\pi\upsilon t}dt$. The spreading function characterizes the reflectivity of scatters associated to delay $\tau$ and Doppler $\upsilon$. For a non-wide-sense stationary uncorrelated scattering (non-WSSUS) channel, we can obtain the following functions $R_S(\tau,\upsilon;\Delta\tau,\Delta\upsilon)=E\{S_{\mathbf{H}}(\tau,\upsilon+\Delta\upsilon)S_{\mathbf{H}}^*(\tau-\Delta\tau,\upsilon)\}$ and $R_L(t,f;\Delta t,\Delta f)=E\{L_{\mathbf{H}}(t,f+\Delta f)L_{\mathbf{H}}^*(t-\Delta t,f)\}$, where $E\{\cdot\}$ denotes the expectation. The received signal can be expressed as $r(t)=(\mathbf{H}x)(t)+w(t)$, where $w(t)$ is the AWGN with variance $\sigma_w^2$.

## 3. Max-SINR Receiver for HMCT Systems

To obtain the data symbol $\hat{c}_{m,n}^i$, the match filter receiver projects the received signal $r(t)$ on prototype pulse function $\psi_{m,n}^i(t)$, $i=1,2$, i.e.,

$$\hat{c}_{m,n}^i=\langle r(t),\psi_{m,n}^i(t)\rangle=\sum_j\sum_{m',n'}c_{m',n'}^j\langle(\mathbf{H}g_{m',n'}^j)(t),\psi_{m,n}^i(t)\rangle+\langle w(t),\psi_{m,n}^i(t)\rangle \qquad (3)$$

where $\langle\cdot\rangle$ denotes the inner product, $\psi_{m,n}^i(t)=\psi(t-mT-\lfloor i/2\rfloor T)e^{-j2\pi(nF+\lfloor i/2\rfloor F)t}$ and $\psi(t)$ is the prototype pulse at the receiver. The energy of the received symbol $\hat{c}_{m,n}^i$ after projection on the prototype pulse set $\psi_{m,n}^i(t)$ can be expressed as

$$E_s=E\left\{\left|\sum_{m',n'}c_{m',n'}^j\langle(\mathbf{H}g_{m',n'}^j)(t),\psi_{m,n}^i(t)\rangle+\langle w(t),\psi_{m,n}^i(t)\rangle\right|^2\right\} \qquad (4)$$

Under the assumption that source symbols are statistically independent, (4) can be rewritten as

$$E_s=\sigma_c^2\iiint R_S(\tau,\upsilon;\Delta\tau,\Delta\upsilon)e^{-j2\pi(f\Delta\tau-t\Delta f\upsilon)}d\Delta\tau d\Delta\upsilon \\ \times\left[\sum_{m,n}\left(\left|A_{g,\psi}(mT+\tau,nF+\upsilon)\right|^2\right)+\left|A_{g,\psi}((m+1/2)T+\tau,(n+1/2)F+\upsilon)\right|^2\right]d\tau d\upsilon+\sigma_w^2\left|A_{g,\psi}(0,0)\right| \qquad (5)$$

The *effective support* $\Xi_\psi$ of the prototype pulse function $\psi(t)$ is approximately the symbol duration $T$. Under the assumptions of doubly underspread channels, that is the local stationarity regions (LSRs) [15,16] of the channel is far greater than $\Xi_\psi$. Hence, we have

$$E\{S_{\mathbf{H}_\kappa}(\tau,\upsilon)S_{\mathbf{H}_\kappa}^*(\tau',\upsilon')\}=C_{\mathbf{H}_\kappa}(\tau,\upsilon)\delta(\tau-\tau')\delta(\upsilon-\upsilon') \qquad (6)$$

and

$$\ell_{\mathbf{H}_\kappa}(t,f;\tau,\upsilon)=\iiint R_S(\tau,\upsilon;\Delta\tau,\Delta\upsilon)e^{-j2\pi(f\Delta\tau-t\Delta f\upsilon)}d\Delta\tau d\Delta\upsilon=C_{\mathbf{H}_\kappa}(\tau,\upsilon) \qquad (7)$$

Substituting (7) in (5), we can rewrite (5) as

$$E_s=\sigma_c^2\int_\tau\int_\upsilon C_{\mathbf{H}_\kappa}(\tau,\upsilon)\left[\sum_{m,n}\left(\left|A_{g,\psi}(mT+\tau,nF+\upsilon)\right|^2\right)+\left|A_{g,\psi}((m+1/2)T+\tau,(n+1/2)F+\upsilon)\right|^2\right]d\tau d\upsilon+\sigma_w^2\left|A_{g,\psi}(0,0)\right| \qquad (8)$$

The SINR of received signal can be expressed as

$$R_{\mathrm{SIN}}^\kappa=\sigma_c^2/E_{\mathrm{IN}}^\kappa\int_\tau\int_\upsilon C_{\mathbf{H}_\kappa}(\tau,\upsilon)\left|A_{g,\psi}(\tau,\upsilon)\right|^2 d\tau d\upsilon \qquad (9)$$

where the interference-plus-noise energy

$$E_{\mathrm{IN}}^\kappa=\sigma_c^2\int_\tau\int_\upsilon C_{\mathbf{H}_\kappa}(\tau,\upsilon)\left[\sum_{z=[m,n]^T\neq[0,0]^T}\left|A_{g,\psi}(mT+\tau,nF+\upsilon)\right|^2 \\ +\sum_{z=[m,n]^T\neq[0,0]^T}\left|A_{g,\psi}(mT+T/2+\tau,nF+F/2+\upsilon)\right|^2\right]d\tau d\upsilon+\sigma_w^2\left|A_{g,\psi}(0,0)\right| \qquad (10)$$

Clearly, the interference-plus-noise energy function $R_{\mathrm{SIN}}^\kappa$ is dependent on the channel scattering function and the pulse shape (through its ambiguity function). For the DD channel with exponential power delay profile and U-shape Doppler spectrum, within a LSR $\kappa$, the scattering function can be expressed as [18]

$$C_{\mathbf{H}}^\kappa(\tau,\upsilon)=e^{-\tau/\tau_{rms}^\kappa}\left(\pi\tau_{rms}^\kappa f_{\max}^\kappa\sqrt{1-(\upsilon/f_{\max}^\kappa)^2}\right)^{-1} \qquad (11)$$

with $\tau\geq 0, |\upsilon|<f_{\max}^\kappa$. $\tau_{rms}^\kappa$ and $f_{\max}^\kappa$ denote the RMS delay spread and the maximum Doppler shift within a LSR $\kappa$, respectively. We assume that $\psi(t)=g(t-\Delta t)$, the theoretical SINR of the received signal over the DD

channel with exponential power delay profile and U-shape Doppler spectrum can be expressed as

$$R_{\text{SIN}}^{\kappa}(\Delta t) = \frac{\sigma_c^2}{\pi \tau_{rms}^{\kappa} f_{\max}^{\kappa} E_{\text{IN}}^{\kappa}} \int_0^{\infty} e^{-\tau/\tau_{rms}^{\kappa}} e^{-\frac{\pi}{\sigma}(\tau - \Delta t)^2} d\tau \int_{-f_{\max}^{\kappa}}^{f_{\max}^{\kappa}} e^{-\sigma \pi (\upsilon)^2} \Big/ \sqrt{1-\left(\upsilon/f_{\max}^{\kappa}\right)^2} \, d\upsilon \tag{12}$$

The theoretical SINR upper bound of the received signal can be expressed as

$$R_{\text{UB}}^{\kappa} = \arg\max_{\Delta t} R_{\text{SIN}}^{\kappa}(\Delta t) \tag{13}$$

Plugging (12) in (13), the Max-SINR prototype pulse can be expressed as $\psi(t) = g(t - \Delta t)$ and (see Appendix)

$$\Delta t = \frac{\sigma}{2\pi \tau_{rms}^{\kappa}} - \frac{1}{3.8592}\sqrt{\frac{\sigma}{2\pi}}\left(3.28\sqrt{\sigma}/\tau_{rms}^{\kappa} - \sqrt{3.28^2 \sigma/\left(\tau_{rms}^{\kappa}\right)^2 - 7.7184\left(\sigma/\left(\tau_{rms}^{\kappa}\right)^2 - 4\right)}\right) \tag{14}$$

We can see from equation (14) that the prototype pulse of the proposed Max-SINR receiver is a function of RMS.

## 4. Simulation and Discussion

In this section, we test the proposed Max-SINR receiver via computer simulations based on the discrete signal model. In the following simulations, the number of subcarriers for HMCT system is chosen to $N$=40, and the length of prototype pulse $N_g$=600. The center carrier frequency is $f_c$=5GHz and the sampling interval is set to $T_s$=10$^{-6}$s. The system parameters of HMCT system are $F$=25kHz, $T$=1×10$^{-4}$s and $\sigma$ is set to $\sigma = T/\sqrt{3}F$. The product $\vartheta = \tau_{rms} f_{\max}$ is referred to as the channel spread factor (CSF). Traditional projection receiver proposed in [7,8] is named as Traditional Projection Receiver (TPR) in the following simulation results. Non-stationary channel is chosen as DD channel with exponential power delay profile and U-shape Doppler spectrum.

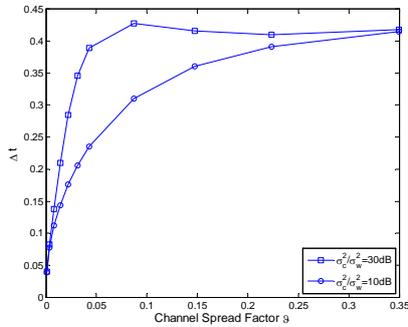 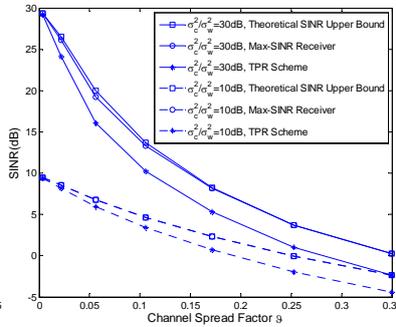 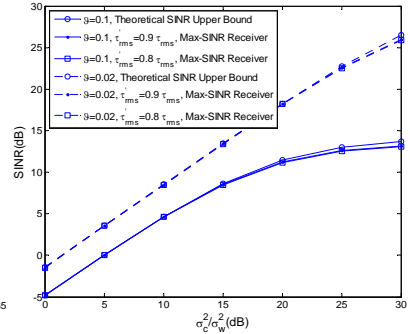

Fig. 1. Delay of the proposed Max-SINR receiver. $\sigma_c^2 / \sigma_w^2$ =10,30dB.

Fig.2. The SINR performance of different receivers. $\sigma_c^2 / \sigma_w^2$ =10,30dB.

Fig. 3. SINR performance of the proposed Max-SINR receiver with estimation error of $\tau_{rms}$.

$\Delta t$ of the proposed Max-SINR receiver with the variety of $\vartheta$ in different LSRs are given in Fig. 1. We can see from Fig. 1 that the delay between the TPR prototype pulse and the Max-SINR prototype pulse increases with the increasing of $\vartheta$ and $\sigma_c^2 / \sigma_w^2$.

The SINR performance of different receivers in LSRs with different $\vartheta$ at $\sigma_c^2 / \sigma_w^2$ =10,30dB are depicted in Fig. 2. The SINR performance of the TPR scheme is depicted for comparison. It can be seen that the SINR performance changes between different LSRs (with different $\vartheta$) and the proposed Max-SINR receiver obtains an approximation to the theoretical upper bound SINR performance within all the LSRs.

The SINR performance of the proposed Max-SINR receiver for HMCT system over DD channel with estimation error of $\tau_{rms}$ is given in Fig. 4. We can conclude from Fig. 4 that the proposed Max-SINR receiver is robust to the estimation error of RMS delay spread.

## 5. Conclusion

The Max-SINR receiver is proposed for HMCT system over non-stationary DD channel in this paper. After a detailed analysis, we present that the prototype pulse of the Max-SINR receiver should adapt to the RMS delay spread within different LSRs. Theoretical analysis shows that the proposed Max-SINR receiver outperforms the TPR scheme in SINR and obtains an approximation to the theoretical upper bound SINR performance within all the LSRs. Simulation results show that the proposed scheme is robust to the estimation error of RMS and is suitable for the actual HMCT system.

## 6. Appendix: Proof of (14)

The SINR of the received signal in (12) can be expressed as

$$R_{\text{SIN}}^{\kappa} \approx \frac{\sigma_c^2}{\sigma_w^2 \tau_{rms}^{\kappa} f_{\max}^{\kappa}} \int_{-f_{\max}^{\kappa}}^{f_{\max}^{\kappa}} e^{-\sigma \pi v^2} \Big/ \sqrt{1-\left(v/f_{\max}^{\kappa}\right)^2}\, dv \left( \underbrace{e^{\frac{\sigma}{4\pi\left(\tau_{rms}^{\kappa}\right)^2}+\frac{\Delta t}{\tau_{rms}^{\kappa}}}}_{a(\Delta t)} \underbrace{\int_0^{\infty} e^{-\frac{\pi}{\sigma}\left(\tau-\Delta t+\frac{\sigma}{2\pi\tau_{rms}^{\kappa}}\right)^2} d\tau}_{b(\Delta t)} \right) \quad (15)$$

where $b(\Delta t)$ in (15) can be rewritten as

$$b(\Delta t) = \sqrt{\frac{\sigma}{\pi}} \int_{\sqrt{\frac{\pi}{\sigma}}\left(\frac{\sigma}{2\pi\tau_{rms}^{\kappa}}-\Delta t\right)}^{\infty} e^{-x^2} dx = \frac{\sqrt{\sigma}}{2} erfc\left(\sqrt{\frac{\pi}{\sigma}}\left(\frac{\sigma}{2\pi\tau_{rms}^{\kappa}}-\Delta t\right)\right) \quad (16)$$

where $erfc(\cdot)$ is the complementary error function. If $x > 0$, we may obtain an approximate solution of the complementary error function $erfc(\cdot)$ by [19]

$$erfc\left(x/\sqrt{2}\right) \approx 2e^{-x^2/2} \Big/ \left(1.64x + \sqrt{0.76x^2 + 4}\right) \quad (17)$$

The Max-SINR receiver can be obtained by solving the gradient of $a(\Delta t)b(\Delta t)$ with respect to $\Delta t$, that is $\left(\partial a(\Delta t)/\partial \Delta t\right)b(\Delta t) = -\left(\partial b(\Delta t)/\partial \Delta t\right)a(\Delta t)$, where $\partial a(\Delta t)/\partial \Delta t = -a(\Delta t)/\tau_{rms}^{\kappa}$, $\partial b(\Delta t)/\partial \Delta t = \exp\left(-\pi/\sigma\left(\sigma/2\pi\tau_{rms}^{\kappa}-\Delta t\right)^2\right)$, hence, we have

$$\begin{aligned}\frac{b(\Delta t)}{\tau_{rms}^{\kappa}} &= e^{-\frac{\pi}{\sigma}\left(\frac{\sigma}{2\pi\tau_{rms}^{\kappa}}-\Delta t\right)^2} = \frac{\sqrt{\sigma}}{2\tau_{rms}^{\kappa}} erfc\left(\sqrt{\frac{\pi}{\sigma}}\left(\frac{\sigma}{2\pi\tau_{rms}^{\kappa}}-\Delta t\right)\right) \\ &\approx \frac{\sqrt{\sigma}}{\tau_{rms}^{\kappa}} e^{-\frac{\pi}{\sigma}\left(\frac{\sigma}{2\pi\tau_{rms}^{\kappa}}-\Delta t\right)^2} \left(1.64\sqrt{\frac{2\pi}{\sigma}}\left(\frac{\sigma}{2\pi\tau_{rms}^{\kappa}}-\Delta t\right) + \sqrt{\frac{1.52\pi}{\sigma}\left(\frac{\sigma}{2\pi\tau_{rms}^{\kappa}}-\Delta t\right)^2 + 4}\right)^{-1}\end{aligned} \quad (18)$$

Equation (18) can be simplified to a quadratic equation. Under the constraint of $\Delta t > 0$, the solution of the quadratic equation can be expressed as equation (14).

## 7. Acknowledgement

This work was supported by the National Natural Science Foundation of China (No. 61301165, 61371123), Jiangsu Province National Science Foundation under Grant (No. BK2012055, BK2011002).